\pdfoutput=1

\documentclass[12pt,a4paper]{article}

\usepackage{ifthen} 
\newboolean{pdflatex}
\setboolean{pdflatex}{true} 

\newboolean{articletitles}
\setboolean{articletitles}{true} 

\newboolean{uprightparticles}
\setboolean{uprightparticles}{false} 

\newboolean{inbibliography}
\setboolean{inbibliography}{false} 

\usepackage{microtype}
\usepackage{lineno}  
\usepackage{xspace} 
\usepackage[tableposition=top]{caption} 

\usepackage{graphicx}  
\usepackage{color}
\usepackage{colortbl}
\graphicspath{{./../Figures/}} 

\usepackage{amsmath} 
\usepackage{amssymb}
\usepackage{amsfonts}
\usepackage{upgreek} 

\newcommand*\patchAmsMathEnvironmentForLineno[1]{%
\expandafter\let\csname old#1\expandafter\endcsname\csname #1\endcsname
\expandafter\let\csname oldend#1\expandafter\endcsname\csname
end#1\endcsname
 \renewenvironment{#1}%
   {\linenomath\csname old#1\endcsname}%
   {\csname oldend#1\endcsname\endlinenomath}%
}
\newcommand*\patchBothAmsMathEnvironmentsForLineno[1]{%
  \patchAmsMathEnvironmentForLineno{#1}%
  \patchAmsMathEnvironmentForLineno{#1*}%
}
\AtBeginDocument{%
\patchBothAmsMathEnvironmentsForLineno{equation}%
\patchBothAmsMathEnvironmentsForLineno{align}%
\patchBothAmsMathEnvironmentsForLineno{flalign}%
\patchBothAmsMathEnvironmentsForLineno{alignat}%
\patchBothAmsMathEnvironmentsForLineno{gather}%
\patchBothAmsMathEnvironmentsForLineno{multline}%
}

\usepackage{hyperref}    
\usepackage[all]{hypcap} 

\def\lhcb {\mbox{LHCb}\xspace}

\ifthenelse{\boolean{uprightparticles}}%
{

 \def\Pmu         {\ensuremath{\upmu}\xspace}

 \def\Ppi         {\ensuremath{\uppi}\xspace}

 \def\Ppsi        {\ensuremath{\uppsi}\xspace}

 \def\PDelta      {\ensuremath{\Delta}\xspace}                 
 \def\PXi      {\ensuremath{\Xi}\xspace}                 
 \def\PLambda      {\ensuremath{\Lambda}\xspace}                 
 \def\PSigma      {\ensuremath{\Sigma}\xspace}                 
 \def\POmega      {\ensuremath{\Omega}\xspace}                 
 \def\PUpsilon      {\ensuremath{\Upsilon}\xspace}                 
 
 \def\PB      {\ensuremath{\mathrm{B}}\xspace}                 
                  
 \def\PD      {\ensuremath{\mathrm{D}}\xspace}

 \def\PJ      {\ensuremath{\mathrm{J}}\xspace}                 
 \def\PK      {\ensuremath{\mathrm{K}}\xspace}

 \def\Pb      {\ensuremath{\mathrm{b}}\xspace}                 
 \def\Pc      {\ensuremath{\mathrm{c}}\xspace}                 
 \def\Pd      {\ensuremath{\mathrm{d}}\xspace}                 
 \def\Pe      {\ensuremath{\mathrm{e}}\xspace}

 \def\Pi      {\ensuremath{\mathrm{i}}\xspace}

 \def\Pp      {\ensuremath{\mathrm{p}}\xspace}                 
 \def\Pq      {\ensuremath{\mathrm{q}}\xspace}                 
                  
 \def\Ps      {\ensuremath{\mathrm{s}}\xspace}

}
{

 \def\Pmu         {\ensuremath{\mu}\xspace}

 \def\Ppi         {\ensuremath{\pi}\xspace}

 \def\Ppsi        {\ensuremath{\psi}\xspace}                 
                  
 \mathchardef\PDelta="7101
 \mathchardef\PXi="7104
 \mathchardef\PLambda="7103
 \mathchardef\PSigma="7106
 \mathchardef\POmega="710A
 \mathchardef\PUpsilon="7107
                  
 \def\PB      {\ensuremath{B}\xspace}                 
                  
 \def\PD      {\ensuremath{D}\xspace}

 \def\PJ      {\ensuremath{J}\xspace}                 
 \def\PK      {\ensuremath{K}\xspace}

 \def\Pb      {\ensuremath{b}\xspace}                 
 \def\Pc      {\ensuremath{c}\xspace}                 
 \def\Pd      {\ensuremath{d}\xspace}                 
 \def\Pe      {\ensuremath{e}\xspace}

 \def\Pi      {\ensuremath{i}\xspace}

 \def\Pp      {\ensuremath{p}\xspace}                 
 \def\Pq      {\ensuremath{q}\xspace}                 
                  
 \def\Ps      {\ensuremath{s}\xspace}

}

\def\ep         {\ensuremath{\Pe^+}\xspace}

\def\mup        {\ensuremath{\Pmu^+}\xspace}

\def\quark     {\ensuremath{\Pq}\xspace}

\def\dquark    {\ensuremath{\Pd}\xspace}

\def\squark    {\ensuremath{\Ps}\xspace}

\def\cquark    {\ensuremath{\Pc}\xspace}
\def\cquarkbar {\ensuremath{\overline \cquark}\xspace}
\def\ccbar     {\ensuremath{\cquark\cquarkbar}\xspace}
\def\bquark    {\ensuremath{\Pb}\xspace}
\def\bquarkbar {\ensuremath{\overline \bquark}\xspace}
\def\bbbar     {\ensuremath{\bquark\bquarkbar}\xspace}

\def\pion  {\ensuremath{\Ppi}\xspace}

\def\pip   {\ensuremath{\pion^+}\xspace}
\def\pim   {\ensuremath{\pion^-}\xspace}

\def\kaon  {\ensuremath{\PK}\xspace}
  \def\Kbar  {\kern 0.2em\overline{\kern -0.2em \PK}{}\xspace}

\def\Kp    {\ensuremath{\kaon^+}\xspace}
\def\Km    {\ensuremath{\kaon^-}\xspace}

\def\Kstarz  {\ensuremath{\kaon^{*0}}\xspace}

  \def\Dbar    {\kern 0.2em\overline{\kern -0.2em \PD}{}\xspace}
\def\D       {\ensuremath{\PD}\xspace}

\def\Dz      {\ensuremath{\D^0}\xspace}

\def\Dp      {\ensuremath{\D^+}\xspace}
\def\Dm      {\ensuremath{\D^-}\xspace}

\def\Dsm     {\ensuremath{\D^-_\squark}\xspace}

\def\B       {\ensuremath{\PB}\xspace}
\def\Bbar    {\ensuremath{\kern 0.18em\overline{\kern -0.18em \PB}{}}\xspace}

\def\Bz      {\ensuremath{\B^0}\xspace}
\def\Bzb     {\ensuremath{\Bbar^0}\xspace}
\def\Bu      {\ensuremath{\B^+}\xspace}
\def\Bub     {\ensuremath{\B^-}\xspace}
\def\Bp      {\ensuremath{\Bu}\xspace}
\def\Bm      {\ensuremath{\Bub}\xspace}

\def\Bd      {\ensuremath{\B^0}\xspace}
\def\Bs      {\ensuremath{\B^0_\squark}\xspace}

\def\Bdb     {\ensuremath{\Bbar^0}\xspace}

\def\jpsi     {\ensuremath{{\PJ\mskip -3mu/\mskip -2mu\Ppsi\mskip 2mu}}\xspace}

  \def\Y#1S{\ensuremath{\PUpsilon{(#1S)}}\xspace}

\def\Lz {\ensuremath{\PLambda}\xspace}
\def\Lbar {\ensuremath{\kern 0.1em\overline{\kern -0.1em\PLambda}}\xspace}

\def\Lc      {\ensuremath{\Lz^+_\cquark}\xspace}

\newcommand{\decay}[2]{\ensuremath{#1\!\to #2}\xspace}         

\def\to                 {\ensuremath{\rightarrow}\xspace}

\def\CP                {\ensuremath{C\!P}\xspace}

\newcommand{\mistag}{\ensuremath{\omega}\xspace}

\newcommand{\etag}{{\ensuremath{\varepsilon_{\rm tag}}}\xspace}

\newcommand{\effeff}{\ensuremath{\varepsilon_{\rm eff}}\xspace}

\def\BsToJPsiPhi  {\decay{\Bs}{\jpsi\phi}}
\def\BdToJPsiKst  {\decay{\Bd}{\jpsi\Kstarz}}

\def\BToJPsiK     {\decay{\B^+}{\jpsi\Kp}}
\def\BdToDpi     {\decay{\Bd}{\Dm\pip}}
\def\BsToDspi     {\decay{\Bs}{\Dsm\pip}}

\def\AT#1     {\ensuremath{A_{\mathrm{T}}^{#1}}\xspace}           

\def\C#1      {\ensuremath{\mathcal{C}_{#1}}\xspace}                       
\def\Cp#1     {\ensuremath{\mathcal{C}_{#1}^{'}}\xspace}                    
\def\Ceff#1   {\ensuremath{\mathcal{C}_{#1}^{\mathrm{(eff)}}}\xspace}        
\def\Cpeff#1  {\ensuremath{\mathcal{C}_{#1}^{'\mathrm{(eff)}}}\xspace}       
\def\Ope#1    {\ensuremath{\mathcal{O}_{#1}}\xspace}                       
\def\Opep#1   {\ensuremath{\mathcal{O}_{#1}^{'}}\xspace}                    

\newcommand{\tev}{\ifthenelse{\boolean{inbibliography}}{\ensuremath{~T\kern -0.05em eV}\xspace}{\ensuremath{\mathrm{\,Te\kern -0.1em V}}\xspace}}
\newcommand{\gev}{\ensuremath{\mathrm{\,Ge\kern -0.1em V}}\xspace}
\newcommand{\mev}{\ensuremath{\mathrm{\,Me\kern -0.1em V}}\xspace}
\newcommand{\kev}{\ensuremath{\mathrm{\,ke\kern -0.1em V}}\xspace}
\newcommand{\ev}{\ensuremath{\mathrm{\,e\kern -0.1em V}}\xspace}
\newcommand{\gevc}{\ensuremath{{\mathrm{\,Ge\kern -0.1em V\!/}c}}\xspace}
\newcommand{\mevc}{\ensuremath{{\mathrm{\,Me\kern -0.1em V\!/}c}}\xspace}
\newcommand{\gevcc}{\ensuremath{{\mathrm{\,Ge\kern -0.1em V\!/}c^2}}\xspace}
\newcommand{\gevgevcccc}{\ensuremath{{\mathrm{\,Ge\kern -0.1em V^2\!/}c^4}}\xspace}
\newcommand{\mevcc}{\ensuremath{{\mathrm{\,Me\kern -0.1em V\!/}c^2}}\xspace}

\def\mum  {\ensuremath{{\,\upmu\rm m}}\xspace}

\def\invfb   {\ensuremath{\mbox{\,fb}^{-1}}\xspace}

\newcommand{\stat}{\ensuremath{\mathrm{\,(stat)}}\xspace}
\newcommand{\syst}{\ensuremath{\mathrm{\,(syst)}}\xspace}

\def\gsim{{~\raise.15em\hbox{$>$}\kern-.85em
          \lower.35em\hbox{$\sim$}~}\xspace}
\def\lsim{{~\raise.15em\hbox{$<$}\kern-.85em
          \lower.35em\hbox{$\sim$}~}\xspace}

\def\sPlot{\mbox{\em sPlot}}

\def\ptot       {\mbox{$p$}\xspace}
\def\pt         {\mbox{$p_{\rm T}$}\xspace}

\def\evtgen     {\mbox{\textsc{EvtGen}}\xspace}

\def\geant      {\mbox{\textsc{Geant4}}\xspace}

\def\photos     {\mbox{\textsc{Photos}}\xspace}

\def\pythia     {\mbox{\textsc{Pythia}}\xspace}

\def\tell1  {TELL1\xspace}
\def\ukl1   {UKL1\xspace}

\usepackage{cite} 
\usepackage{mciteplus}

\usepackage{multicol}
\usepackage[capitalize]{cleveref}

\def\Let@{\def\\{\notag\math@cr}}

\usepackage{multirow}

\usepackage{siunitx}
\usepackage{booktabs}

\newcommand{\avg}[1]{\langle #1 \rangle}

\newcommand{\modeZero}{$\Dz \to \Km \pip $}
\newcommand{\modeOne}{$\Dz \to \Km \pip \pip \pim$}
\newcommand{\modeTwo}{$\Dp \to \Km \pip \pip$}
\newcommand{\modeThree}{$H_c \to \Km \pip X$}
\newcommand{\modeFour}{$H_c \to \Km \ep X$}
\newcommand{\modeFive}{$H_c \to \Km \mup X$}
\newcommand{\modeSix}{$\Lc \to \Pp^+ \Km \pip$}

\newcommand{\PowerMCShort}{$0.33\%$}

\newcommand{\RateMCShort}{$4.88\%$}

\newcommand{\MistagMCShort}{$37.0\%$}

\newcommand{\PowerXAllShort}{$(0.30 \pm 0.01 \pm 0.01) \%$}

\newcommand{\PowerXAllLabelShort}{$(0.30 \pm 0.01 \text{\stat} \pm 0.01 \text{\syst}) \%$}

\newcommand{\RateXAllShort}{$(3.11 \pm 0.02) \%$}

\newcommand{\MistagXAllShort}{$(34.6 \pm 0.3 \pm 0.3) \%$}

\newcommand{\KstPowerXAllShort}{$(0.30 \pm 0.03 \pm 0.01) \%$}

\newcommand{\KstRateXAllShort}{$(3.32 \pm 0.04) \%$}

\newcommand{\KstMistagXAllShort}{$(35.0 \pm 0.8 \pm 0.3) \%$}

\newcommand{\DPowerXAllShort}{$(0.40 \pm 0.02 \pm 0.01) \%$}

\newcommand{\DPowerXAllLabelShort}{$(0.40 \pm 0.02 \text{\stat} \pm 0.01 \text{\syst}) \%$}

\newcommand{\DRateXAllShort}{$(4.11 \pm 0.03) \%$}

\newcommand{\DMistagXAllShort}{$(34.4 \pm 0.4 \pm 0.3) \%$}

\newcommand{\DsPowerXAllShort}{$(0.39 \pm 0.03 \pm 0.01) \%$}

\newcommand{\DsPowerXAllLabelShort}{$(0.39 \pm 0.03 \text{\stat} \pm 0.01 \text{\syst}) \%$}

\newcommand{\DsRateXAllShort}{$(3.99 \pm 0.07) \%$}

\newcommand{\DsMistagXAllShort}{$(34.4 \pm 0.6 \pm 0.3) \%$}

\newcommand{\numDataCandidatesTotal}{$1.1 \times 10^6$} 

\begin{document}

\renewcommand{\thefootnote}{\fnsymbol{footnote}}
\setcounter{footnote}{1}

\begin{titlepage}
\pagenumbering{roman}

\vspace*{-1.5cm}
\centerline{\large EUROPEAN ORGANIZATION FOR NUCLEAR RESEARCH (CERN)}
\vspace*{1.5cm}
\noindent
\begin{tabular*}{\linewidth}{lc@{\extracolsep{\fill}}r@{\extracolsep{0pt}}}
\ifthenelse{\boolean{pdflatex}}
{\vspace*{-2.7cm}\mbox{\!\!\!\includegraphics[width=.14\textwidth]{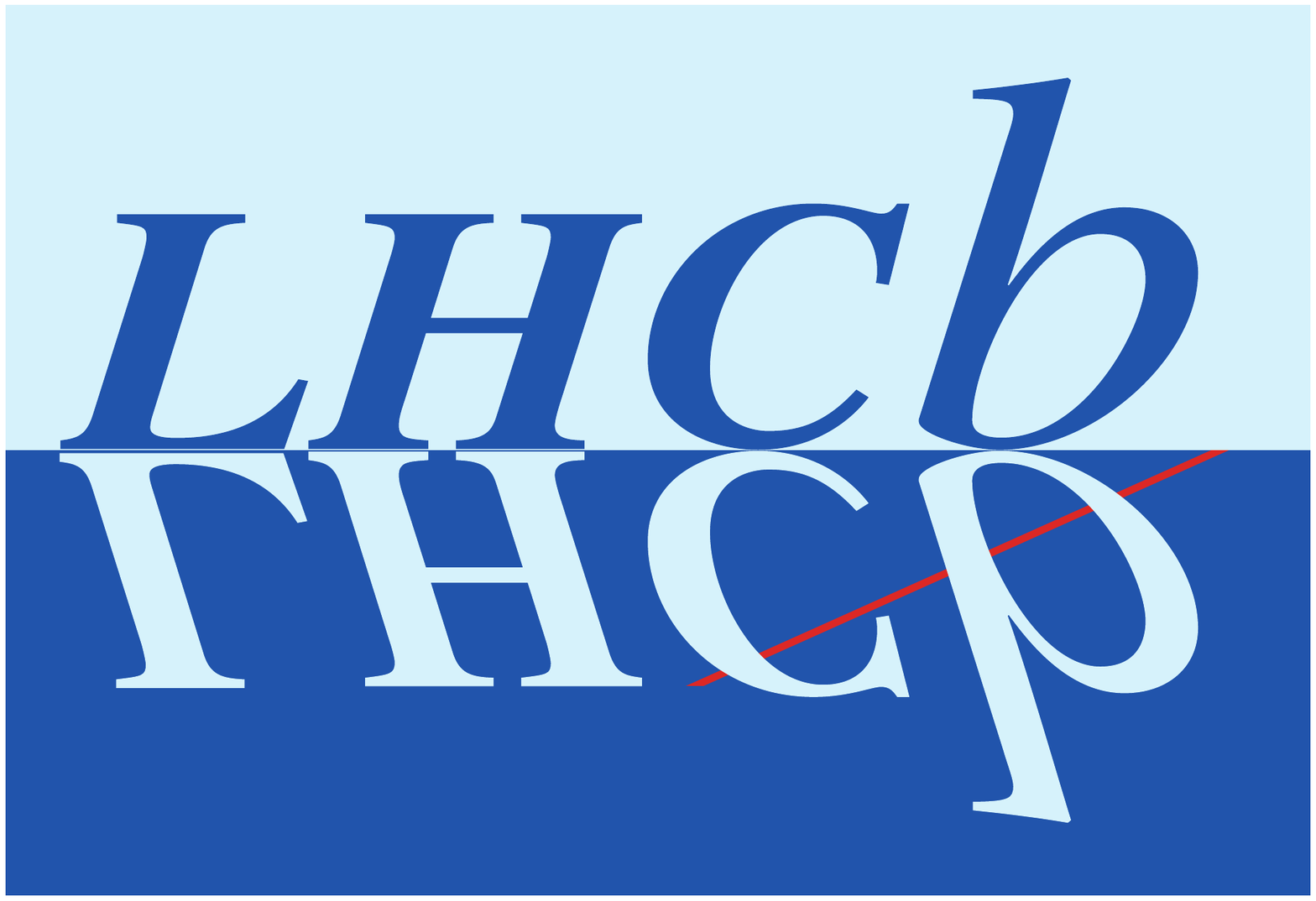}} & &}%
{\vspace*{-1.2cm}\mbox{\!\!\!\includegraphics[width=.12\textwidth]{lhcb-logo.eps}} & &}%
\\
 & & CERN-PH-EP-2015-193 \\  
 & & LHCb-PAPER-2015-027 \\  
 & & 5 October 2015 \\ 
 & & \\
\end{tabular*}

\vspace*{3.0cm}

{\bf\boldmath\LARGE
\begin{center}
  \B flavour tagging using charm decays at the \lhcb experiment
\end{center}
}

\vspace*{2.0cm}

\begin{center}
The LHCb collaboration\footnote{Authors are listed at the end of this paper.}
\end{center}

\vspace{\fill}

\begin{abstract}
  \noindent
 An algorithm is described for tagging the flavour content at production of neutral \B mesons in the \lhcb experiment.
 The algorithm exploits the correlation of the flavour of a \B meson with the charge of a reconstructed secondary charm hadron from the decay of the  other \bquark hadron produced in the proton-proton collision.
 Charm hadron candidates are identified in a number of fully or partially reconstructed Cabibbo-favoured decay modes.
  The algorithm is calibrated on the self-tagged decay modes \BToJPsiK and \BdToJPsiKst using $3.0\invfb$ of data collected by the \lhcb experiment at $pp$ centre-of-mass energies of $7\tev$ and $8\tev$.
  Its tagging power on these samples of $\B \to \jpsi X$ decays is \PowerXAllShort{}. 
\end{abstract}

\vspace*{1.0cm}

\begin{center}
  Submitted to J.~Instrum.
\end{center}

\vspace{\fill}

{\footnotesize 
\centerline{\copyright~CERN on behalf of the \lhcb collaboration, licence \href{http://creativecommons.org/licenses/by/4.0/}{CC-BY-4.0}.}}
\vspace*{2mm}

\end{titlepage}


\newpage
\setcounter{page}{2}
\mbox{~}
%

\cleardoublepage


\renewcommand{\thefootnote}{\arabic{footnote}}
\setcounter{footnote}{0}


\pagestyle{plain} 
\setcounter{page}{1}
\pagenumbering{arabic}


\section{Introduction}
\label{sec:intro}

Measurements that involve mixing and time-dependent \CP asymmetries in decays of neutral \B mesons require the identification of their flavour content at production.
This is achieved via various flavour tagging algorithms that exploit information from the rest of the $pp$ collision event.
Same-side (SS) taggers look for particles produced in association with the signal \B meson during the hadronization of the \bquark quark~\cite{Gronau:1992ke}.
The \dquark or \squark partner of the light valence quark of the signal \B has a roughly $50\%$ chance of hadronizing into a charged pion or kaon.
Since \bquark quarks are mostly produced in \bbbar pairs, the flavour content of the signal \B meson can also be deduced from available information on the opposite-side (OS) \bquark hadron, whose flavour is the opposite of the signal \B meson at the production time.
OS muon and electron taggers look for leptons originating from semileptonic $\bquark \to \cquark W$ transitions of the \bquark hadron,
and an OS kaon tagger looks for kaons coming from $\bquark \to \cquark \to \squark$ transitions. 
A vertex-charge tagger reconstructs the decay vertex of the OS \bquark hadron and predicts its charge by weighting the charges of its decay products according to their transverse momentum.
The OS taggers employed by LHCb are described in Ref.~\cite{LHCb-PAPER-2011-027} and the SS taggers in Refs.~\cite{LHCb-PAPER-2015-004,LHCb-CONF-2012-033}.
This paper reports a new flavour tagging algorithm for the \lhcb experiment that relies on reconstructed decays of charm hadrons produced in the OS \bquark hadron decay.
For the development and evaluation of the tagging algorithm, signal \B meson and charm hadron candidates are reconstructed using data from 3\,fb$^{-1}$ of integrated luminosity collected by \lhcb at $7\tev$ and $8\tev$ centre-of-mass energies in 2011 and 2012, respectively.

The performance of a flavour tagging algorithm is defined by its tagging efficiency, \etag, mistag fraction, \mistag, and dilution, $\mathcal{D}=1-2\mistag$.  
For a simple tagging algorithm with discrete decisions -- $\Bz$, $\Bzb$, or untagged -- these metrics are directly related to the 
numbers of rightly tagged ($R$), wrongly tagged ($W$), and untagged events ($U$) in a signal sample:
\begin{equation}
	\etag = \frac{R+W}{R+W+U}, \quad
	\mistag = \frac{W}{R+W}, \quad
	\mathcal{D} = \frac{R-W}{R+W}.
\end{equation}
The performance of the flavour tagging algorithms is improved by assigning confidence weights to their tagging decisions. 
For each tagger, a multivariate classifier is trained using simulated data to distinguish between correct and incorrect decisions ~\cite{LHCb-PAPER-2011-027}.
The inputs to the classifier are a selection of kinematic and geometric quantities describing the tagging track(s), the signal \B meson, and the event.
This classifier then calculates a predicted mistag probability $\eta$ for each decision made. The predicted mistag probability is calibrated to data using an appropriate flavour self-tagged mode, 
such as \BToJPsiK, or a mode involving neutral \B oscillation, which self-tags its flavour at the decay-time,  such as \BdToJPsiKst or \BsToDspi ~\cite{LHCb-CONF-2012-026,LHCb-CONF-2012-033} (the use of charge-conjugate modes is implied throughout this paper).
This calibration procedure provides a function $\omega(\eta)$, which relates the actual mistag probability $\omega$ to the predicted mistag probability $\eta$.
Weighting each signal candidate by $1-2\omega(\eta)$ leads to an improved effective mistag fraction $\mistag$ and associated dilution $\mathcal{D} = 1 - 2 \mistag$.
The statistical power of a \CP asymmetry measurement using a tagging algorithm is proportional to the effective tagging efficiency (or tagging power) \effeff, defined as
\begin{equation}
    \effeff = \etag \mathcal{D}^2.
	\label{eq:taggingPower}
\end{equation}
The typical combined tagging power of the current set of OS tagging algorithms used by LHCb is approximately $2.5\%$~\cite{LHCb-PAPER-2014-058,LHCb-PAPER-2014-059,LHCb-PAPER-2015-004,LHCb-PAPER-2015-005}.
Any augmentation to this tagging power increases the statistical precision achievable in $\CP$ measurements at LHCb.

\section{Detector and simulation}

The \lhcb detector~\cite{Alves:2008zz,LHCb-DP-2014-002} is a single-arm forward
spectrometer covering the \mbox{pseudorapidity} range between $2$ and $5$,
designed for the study of particles containing \bquark or \cquark
quarks. The detector includes a high-precision tracking system
consisting of a silicon-strip vertex detector surrounding the $pp$
interaction region~\cite{LHCb-DP-2014-001}, a large-area silicon-strip detector located
upstream of a dipole magnet with a bending power of about
$4{\rm\,Tm}$, and three stations of silicon-strip detectors and straw
drift tubes~\cite{LHCb-DP-2013-003} placed downstream of the magnet.
The tracking system provides a measurement of momentum, \ptot, of charged particles with
a relative uncertainty that varies from 0.5\% at low momentum to 1.0\% at 200\gevc.
The minimum distance of a track to a primary vertex (PV), the impact parameter, is measured with a resolution of $(15+29/\pt)\mum$,
where \pt is the component of the momentum transverse to the beam, in\,\gevc.
Different types of charged hadrons
in the momentum range $2\text{--}100 \, \gevc$
are distinguished using information
from two ring-imaging Cherenkov detectors~\cite{LHCb-DP-2012-003}. 
Photons, electrons and hadrons are identified by a calorimeter system consisting of
scintillating-pad and preshower detectors, an electromagnetic
calorimeter and a hadronic calorimeter. Muons are identified by a
system composed of alternating layers of iron and multiwire
proportional chambers~\cite{LHCb-DP-2012-002}.
The online event selection is performed by a trigger~\cite{LHCb-DP-2012-004}, 
which consists of a hardware stage, based on information from the calorimeter and muon
systems, followed by a software stage, which applies a full event
reconstruction.

In the simulation, $pp$ collisions are generated using
\pythia~\cite{Sjostrand:2006za,Sjostrand:2007gs} 
 with a specific \lhcb
configuration~\cite{LHCb-PROC-2010-056}.  Decays of hadronic particles
are described by \evtgen~\cite{Lange:2001uf}, in which final-state
radiation is generated using \photos~\cite{Golonka:2005pn}. The
interaction of the generated particles with the detector, and its response,
are implemented using the \geant
toolkit~\cite{Allison:2006ve, *Agostinelli:2002hh} as described in
Ref.~\cite{LHCb-PROC-2011-006}.

\section{Tagging potential of OS charm hadrons}
\label{sec:theoryCharm}

In events containing a signal \B decay, the opposite-side \Dp, \Dz, and \Lc charm hadrons are primarily produced through the quark-level $\bquark \to \cquark$ transition.
The charge of the \Dp or \Lc  determines the flavour of the  \bquark hadron parent.
For \Dz decays through the dominant Cabibbo-favoured process $\Dz \to \Km X$, the kaon charge determines the flavour of the charm hadron, and thereby that of the parent \B hadron (the effect of \Dz mixing is negligible).
The OS charm tagging algorithm uses charm hadron candidates reconstructed in a number of decay modes,  chosen for their relatively large branching fractions, listed in Table~\ref{table:decay-modes}.
These include fully reconstructed hadronic modes with a single charged kaon in the final state, partially reconstructed hadronic modes with an unobserved neutral pion, and partially reconstructed semileptonic modes.
Table~\ref{table:decay-modes} also reports the breakdown of the charm tagger's performance by decay mode.
The relative rate and relative power of each mode are the amounts that it contributes to the algorithm's total tagging rate $\etag$ and tagging power $\effeff$, which are presented in Section~\ref{sec:results} and Table~\ref{table:power}.
The algorithm predicts the flavour of the signal \B meson using the charge of the kaon in the same manner as the OS kaon tagger; however, the selection based on the reconstruction of \cquark hadrons (rather than the selection of kaons based on their individual kinematic properties) results in a different set of selected kaons and provides a complementary source of tagging information.

Several effects contribute an irreducible component to the mistag probability for the OS charm tagging algorithm.
The dominant impact comes from
$\Bz$--$\Bzb$ oscillation and from the contributions of ``wrong sign'' charm hadrons produced in $\bquark \to \ccbar \quark$ transitions.
The impact of Cabibbo-suppressed $\Dz \to \Kp X$ decays is negligible, as these typically produce additional kaons and do not mimic modes used by the tagging algorithm,
and doubly Cabibbo-suppressed decays such as $\Dz \to \Kp \pim$ have a negligibly small branching fraction. 
Accounting for relative production rates of \bquark hadrons, neutral \B oscillation, and branching fractions of the decay modes used in the tagger, the irreducible mistag probabilities for \Dz, \Dp and \Lc modes are estimated to be $23 \%$, $19 \%$, and $6 \%$, respectively.

In addition to the irreducible mistag probability arising from physics effects, the charm hadron candidates are contaminated by combinatorial
and partially reconstructed \bquark and \cquark hadron background that can lead to an incorrect flavour tag result.
For each mode, the charm tagger uses a multivariate algorithm that combines geometric and kinematic quantities and properties of the \cquark hadron candidate and its daughters. The resulting discriminating variable is used both to suppress the combinatorial background and to predict the corresponding mistag probability for the surviving candidate.

\section{Selection of charm candidates}
\label{sec:charmmodes}

Charm decay candidates are formed by combining kaon, pion, and proton candidates that satisfy particle identification criteria.
These particles are required to have momentum $p>1000 \mevc$, transverse momentum with respect to the beam axis $p_\mathrm{T} >100 \mevc$, and to be significantly displaced from any PV.
For the candidates in the partially reconstructed modes and the
decay \modeOne{}, which contain large combinatorial backgrounds, more stringent requirements are imposed on the displacement of the final-state particles from the PV.
In addition, particles are required to have $p_\mathrm{T} >150 \mevc$ for candidates in the mode \modeOne{}.

\begin{table}
\caption{Decay modes used in the OS charm tagger. The symbol $H_c$ stands for any \cquark hadron. The definition of the two right-most columns is given in the text.}
\label{table:decay-modes}
\centering
\begin{tabular}{lrr}
\toprule
Decay mode & Relative rate & Relative power \\
\hline
\modeZero{}  & 10.0\% & 24.0\%  \\
\modeOne{}   & 5.9\%  & 8.4\%  \\
\modeTwo{}   & 10.3\% & 2.6\%  \\
\modeThree{} & 69.7\% & 61.5\% \\
\modeFour{}  & 0.5\%  & 0.2\%  \\
\modeFive{}  & 3.4\%  & 0.3\%  \\ 
\modeSix{}   & 0.2\%  & 2.4\%  \\
\bottomrule
\end{tabular}
\end{table}

Charm hadron candidates are required to pass a number of selection requirements.
These include a maximum distance of closest approach between each pair of daughter tracks and a minimum quality of the decay vertex fit.
Each candidate is required to be well separated from any PV and to have a trajectory that leads back to the best PV, chosen to be the PV for which the impact parameter significance of the charm hadron is smallest.
The invariant mass of the charm hadron candidate is required to be consistent with the known mass of the corresponding charm hadron, 
within $100 \mevcc$ for the \Lc channel and $50 \mevcc$ for all other fully reconstructed \D decay modes. 
For the partially reconstructed $D\to K^-\pi^+X$ modes, the $K^-\pi^+$ mass is required to be in a $\left[-400 \mevcc, +0\mevcc \right]$ window around the known \Dz mass 
or in a window of $\pm 50 \mevcc$ around the $K^*(892)^0$ resonance. The former is favoured by the invariant mass distribution of $\Km \pip$ pairs from the quasi-two body decay $\Dz \to \Km \rho^+$, and the latter selects  $\D \to K^*(892)^0 X$ decays.
Charm candidates surviving these criteria still contain significant background contamination, which must be further reduced in order to lower the mistag probability of the algorithm.

For each mode, an adaptive-boosted decision tree (BDT)~\cite{Breiman,AdaBoost} is used both to suppress background candidates and to estimate mistag probabilities.
The inputs to the BDT are variables describing the decay kinematics, decay vertex and displacement, and particle identification information on the decay products.
A variable related to the decay-time is calculated from the distance between the \cquark hadron's decay vertex and the corresponding best PV; this approximates the sum of the decay-times of the \cquark hadron and its parent \bquark hadron.
The BDT algorithms are trained using Monte Carlo (MC) simulations of \bbbar events containing  \BToJPsiK, \BdToJPsiKst, and \BsToJPsiPhi decays on the signal side and inclusive decays of the \bquark hadron on the opposite-side.
These \B decays are used to model the various sources and amounts of background when reconstructing OS \cquark hadrons recoiling against signal \B decays.

The output of the BDT, along with the simulation record of candidate identification, is used to compute the predicted mistag probability $\eta$ for each \cquark hadron candidate. 
Candidates with $\eta < 45\%$ are used in the flavour tagging decision.  Removing candidates that fail this criterion significantly reduces the computing time of the algorithm at little cost to tagging performance.
In cases where multiple charm candidates are present, the candidate with the lowest predicted mistag probability is retained.
The combined efficiency of these requirements for retaining tagged events is $(59.00 \pm 0.07)\%$ and $(53.4\pm 0.3)\%$ in simulation and data, respectively.

\section{Calibration}
\label{sec:calibration}

While simulated data are used to develop and optimize the charm tagging algorithm, its performance is calibrated with collision data by comparing the algorithm's predictions to the known flavours of signal \B candidates, according to the procedure detailed in Ref.~\cite{LHCb-PAPER-2011-027}.
The calibration parameters $p_0$, $p_1$, $\Delta p_0$, and $\Delta p_1$ are defined by
\begin{align*}
\omega &= p_0 + p_1 \left( \eta - \langle \eta \rangle \right) \\
\Delta \omega &= \Delta p_0 + \Delta p_1 \left( \eta - \langle \eta \rangle \right)
\end{align*}
where $\avg{\eta}$ is the average predicted mistag probability, $\omega$ is the actual mistag probability averaged over \Bp and \Bm signal mesons,  and $\Delta \omega$ is the excess mistag probability for \Bp mesons with respect to \Bm mesons; equivalent definitions hold for \Bz/\Bzb signal.
In the ideal case, the offset parameter $p_0$ should equal $\avg{\eta}$, and so the related parameter $\delta p_0 = p_0 - \avg{\eta}$ is often more convenient.

\begin{figure}
\centering
\ifthenelse{\boolean{pdflatex}}
{\includegraphics[scale=0.5]{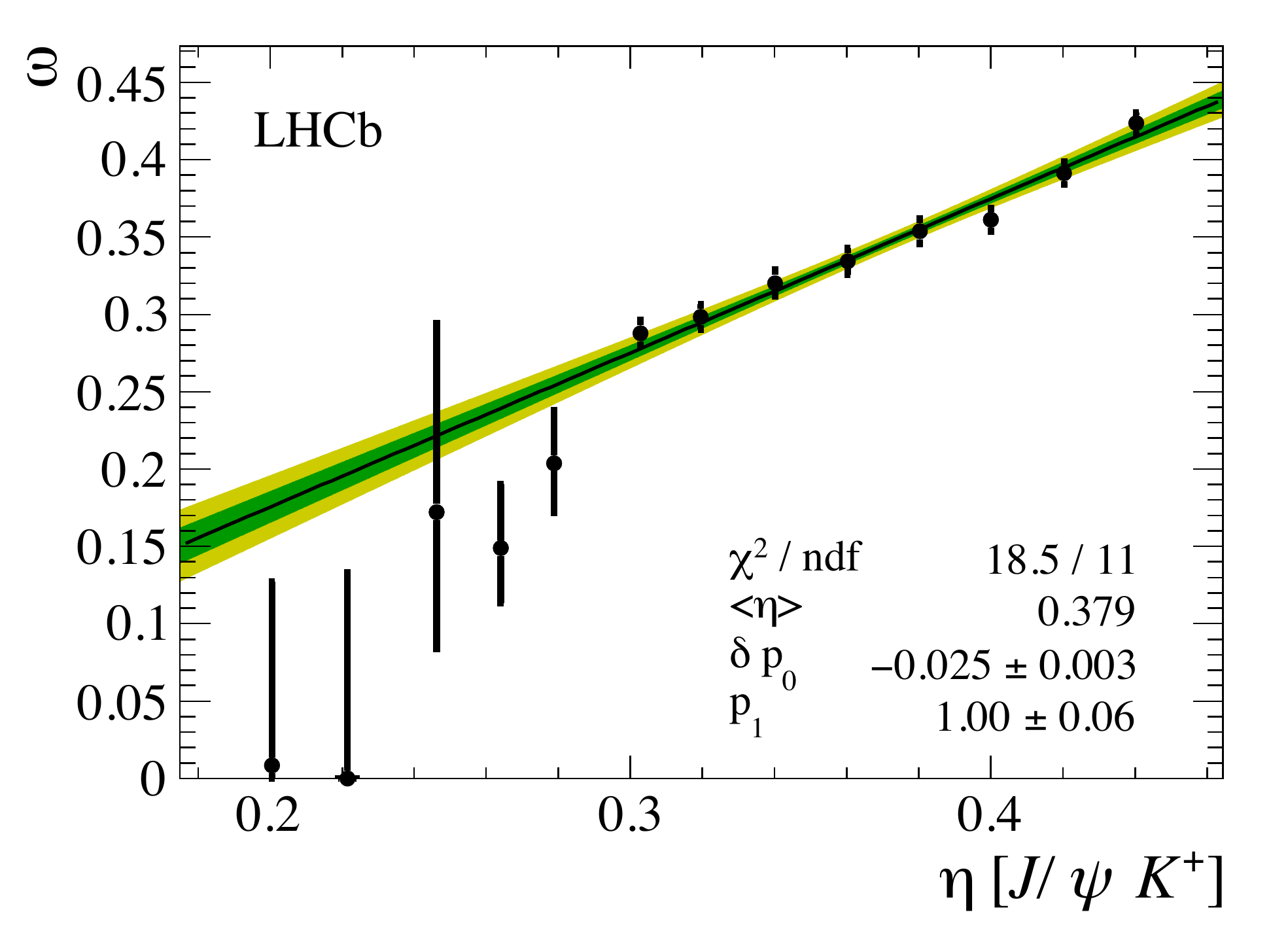}}
{\includegraphics[scale=0.5]{calfit.eps}}
     \caption{Mistag probability $\omega$ as a function of the predicted mistag probability $\eta$ for the \BToJPsiK data sample. A straight line fit to extract the parameters $p_0$ and $p_1$ is superimposed. The dark (green) and light (yellow) bands are the regions within $1\sigma$ and $2\sigma$ of the fitted value, respectively.}
     \label{fig:Cal_All}
\end{figure}

\begin{figure}
\centering
\ifthenelse{\boolean{pdflatex}}
{\includegraphics[scale=0.5]{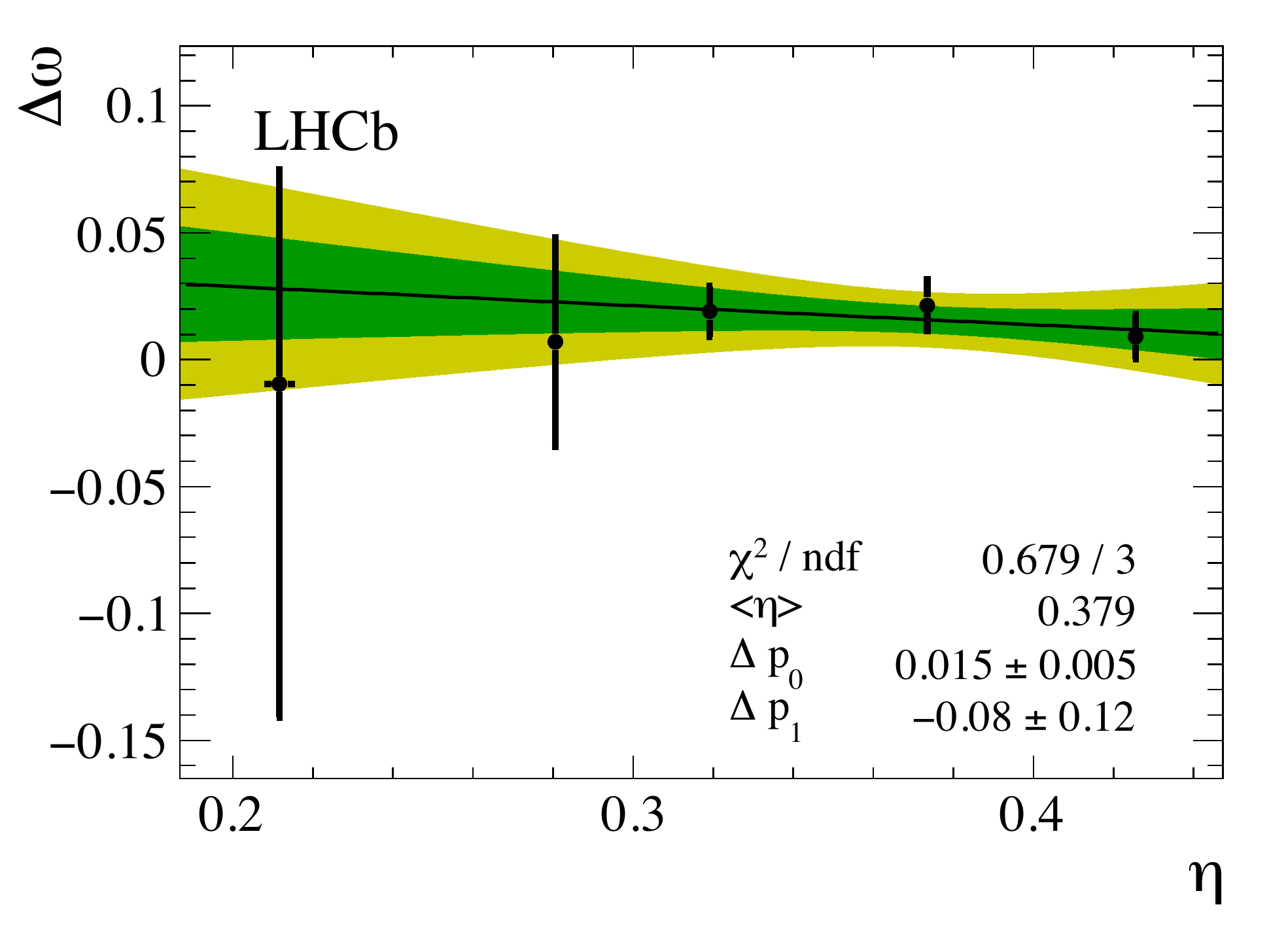}}
{\includegraphics[scale=0.5]{deltacalfit.eps}}
     \caption{Excess mistag probability $\Delta \omega$ as a function of the predicted mistag probability $\eta$ for the \BToJPsiK data sample. A straight line fit to extract the parameters $\Delta p_0$ and $\Delta p_1$ is superimposed. The dark (green) and light (yellow) bands are the regions within $1\sigma$ and $2\sigma$ of the fitted value, respectively.}
     \label{fig:DeltaCal_All}
\end{figure}

A calibration of the algorithm has been performed using the flavour self-tagged mode $\Bp \to \jpsi \Kp$.
The signal candidates are selected by combining pairs of oppositely charged muons, with invariant mass consistent with the known \jpsi mass, with charged kaons, and are required to pass a set of cuts to obtain a good signal to background ratio \cite{LHCb-PAPER-2011-027}.
When multiple candidates are present for a single event, that with the best decay vertex fit is kept.
A fit to the reconstructed \Bp mass distribution is used to separate signal and background via the \sPlot{} procedure, which computes signal and background weights for each candidate~\cite{Pivk:2004ty}.
The empirical model for the signal is a sum of two Crystal Ball functions \cite{Skwarnicki:1986xj}, while background is modeled by an exponential distribution. 
A total of \numDataCandidatesTotal{} signal candidates in this channel are found in the full dataset. 
The parameters $p_0$ and $p_1$ are determined by splitting the data into 13 bins of $\eta$ between 0.19 and 0.45, calculating $\omega_i$ and $\bar{\eta}_i$ (the average $\eta$) in each bin, and performing a linear fit to the set of values $(\omega_i,\bar{\eta}_i)$.
The calibration parameters $\Delta p_0$ and $\Delta p_1$ are obtained from fits to the \Bp and \Bm data each split into 5 bins of $\eta$.
The quantities $\Delta \omega_i$ and $\bar{\eta}_i$ are calculated in each of the 5 bins, and a linear fit is performed to the set of values $(\Delta \omega_i, \bar{\eta}_i)$.
These fits are shown in Figs.~\ref{fig:Cal_All} and \ref{fig:DeltaCal_All}.
The resulting calibration parameters are given in Table~\ref{table:bjpsik-calibration}.

\begin{table}
\caption{Calibration parameters as determined from the \BToJPsiK and \BdToJPsiKst control samples. For both calibration modes, the average predicted mistag probability $\avg{\eta}$ is $0.379$. The first uncertainties are statistical and the second are systematic. The systematic uncertainties are evaluated using simulation.}
\label{table:bjpsik-calibration}
\centering
\begin{tabular}{l l l l l }
\toprule
Sample & \multicolumn{1}{c}{$\phantom{-}\delta p_0$ ($10^{-3}$)} & \multicolumn{1}{c}{$p_1$} & \multicolumn{1}{c}{$\Delta p_0$ ($10^{-3}$)} & \multicolumn{1}{c}{$\Delta p_1$} \\
\hline
\BToJPsiK & $-25\pm3\pm3$ & $1.00\pm0.06\pm0.02$ & $15\pm5\pm4$ & $-0.08\pm0.12\pm0.04$ \\
\BdToJPsiKst & $-18\pm8\pm3$ & $1.16\pm0.17\pm0.02$ & $23\pm11\pm4$ & $\phantom{-}0.21\pm0.25\pm0.04$ \\
\bottomrule
\end{tabular}
\end{table}

\begin{figure}
\centering
\ifthenelse{\boolean{pdflatex}}
{\includegraphics[scale=0.5]{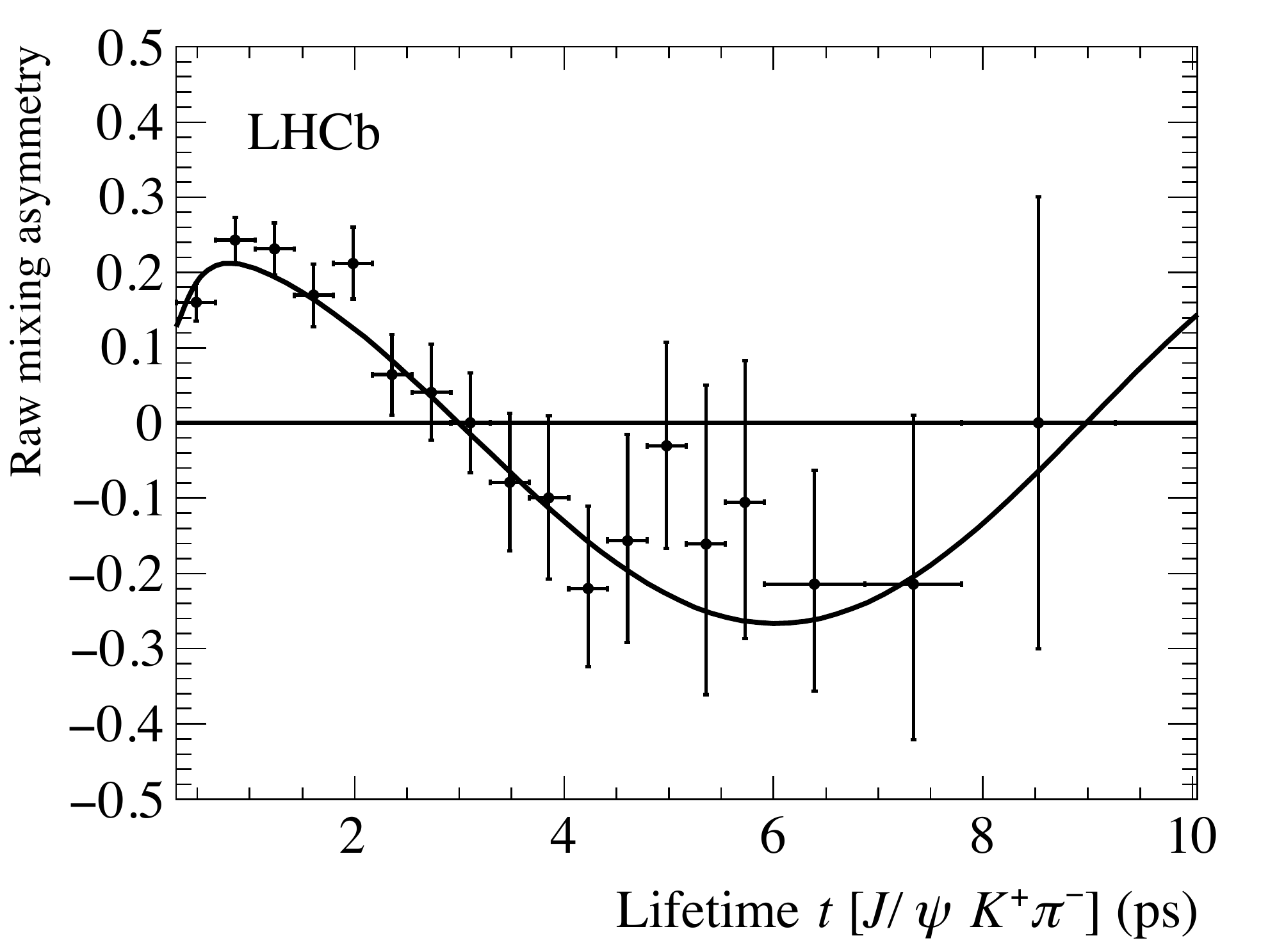}}
{\includegraphics[scale=0.5]{asymmetry.eps}}
     \caption{Raw \Bd--\Bdb mixing asymmetry (defined in Eq.~\protect\ref{eq:mixing-asymmetry}) vs. decay-time for the \BdToJPsiKst data sample. The amplitude of the asymmetry is diluted due to mistagging by the charm tagger. The mixing asymmetry from the fit is superimposed. }
     \label{fig:JpsiKstAsymm}
\end{figure}

A cross-check of the calibration has been carried out using a \BdToJPsiKst control sample.
For this calibration, \Bd--\Bdb oscillation must be taken into account.
The Hypatia function~\cite{Santos:2013gra} is used to model the signal's mass distribution, while the background is modeled with a sum of two exponential functions.
A set of simultaneous fits to the \Bd lifetime distribution in bins of $\eta$ is performed, in which $p_0$, $p_1$, $\Delta p_0$, and $\Delta p_1$ are parameters of the fit model.
In each bin, the raw \Bd--\Bdb mixing asymmetry is defined as
\begin{equation}
	\label{eq:mixing-asymmetry}
	\mathcal{A}_\text{mixing} = \frac{\mathcal{N}(\mathcal{D}=\mathcal{P})-\mathcal{N}(\mathcal{D} \neq \mathcal{P})}{\mathcal{N}(\mathcal{D}=\mathcal{P})+\mathcal{N}(\mathcal{D}\neq \mathcal{P})},
\end{equation}
where $\mathcal{D}$ is the \B meson flavour at decay-time and $\mathcal{P}$ is the production flavour predicted by the charm tagger.
The amplitude of this asymmetry is governed by the actual mistag fraction $\omega_i$ in the bin, while the bin's average predicted mistag probability is $\bar \eta_i$.
The fit attempts to match the calibrated value $\omega(\bar \eta_i)$ to $\omega_i$ in each bin by adjusting the calibration parameters.
A projection of the fitted model to the mixing asymmetry is shown in Fig.~\ref{fig:JpsiKstAsymm}.
The values of the calibration parameters obtained from the fit are given in Table~\ref{table:bjpsik-calibration}.
The parameters are compatible with those obtained in the \BToJPsiK mode, with the total $\chi^2$ per degree of freedom equal to $0.65$.

The relatively small yield of the decay \BsToDspi precludes performing a data-driven calibration on a \Bs mode.
Therefore, in order to ensure that the algorithm performs similarly for \Bs channels as well as \Bp and \Bd channels, separate calibrations to simulated \BToJPsiK, \BdToJPsiKst, and \BsToJPsiPhi events are performed.
Where statistically significant differences between the calibration parameters in the three channels are found, a systematic uncertainty, corresponding to half of the maximum difference, is assigned to the parameter.
These systematic uncertainties are roughly the size of the statistical uncertainties for the parameters $p_0$ and $\Delta p_0$, but are negligible for $p_1$ and $\Delta p_1$.
The propagation of these uncertainties results in a $0.011\%$ absolute systematic uncertainty on the tagging power, comparable to its statistical uncertainty.

Other sources of systematic uncertainty  on calibration parameters have been investigated and found to have negligible effect.
These include the potential effect of the chosen model of the invariant \B mass distribution for the channel \BToJPsiK.
Two alternative models of the mass distribution were used and gave nearly identical results.

There are additional systematic uncertainties related to flavour tagging that must be considered in a $C\!P$ asymmetry analysis.
These include differences between the signal channel sample and calibration channel sample in phase space distribution, event multiplicity, number of primary vertices, or other variables.
These differences would require corrections and would introduce tagging-related systematic uncertainties.
Such effects are dependent on the signal channel and selection, and must be determined separately for each analysis.

\section{Performance}
\label{sec:results}

\begin{figure}
\centering
\ifthenelse{\boolean{pdflatex}}
{\includegraphics[scale=0.5]{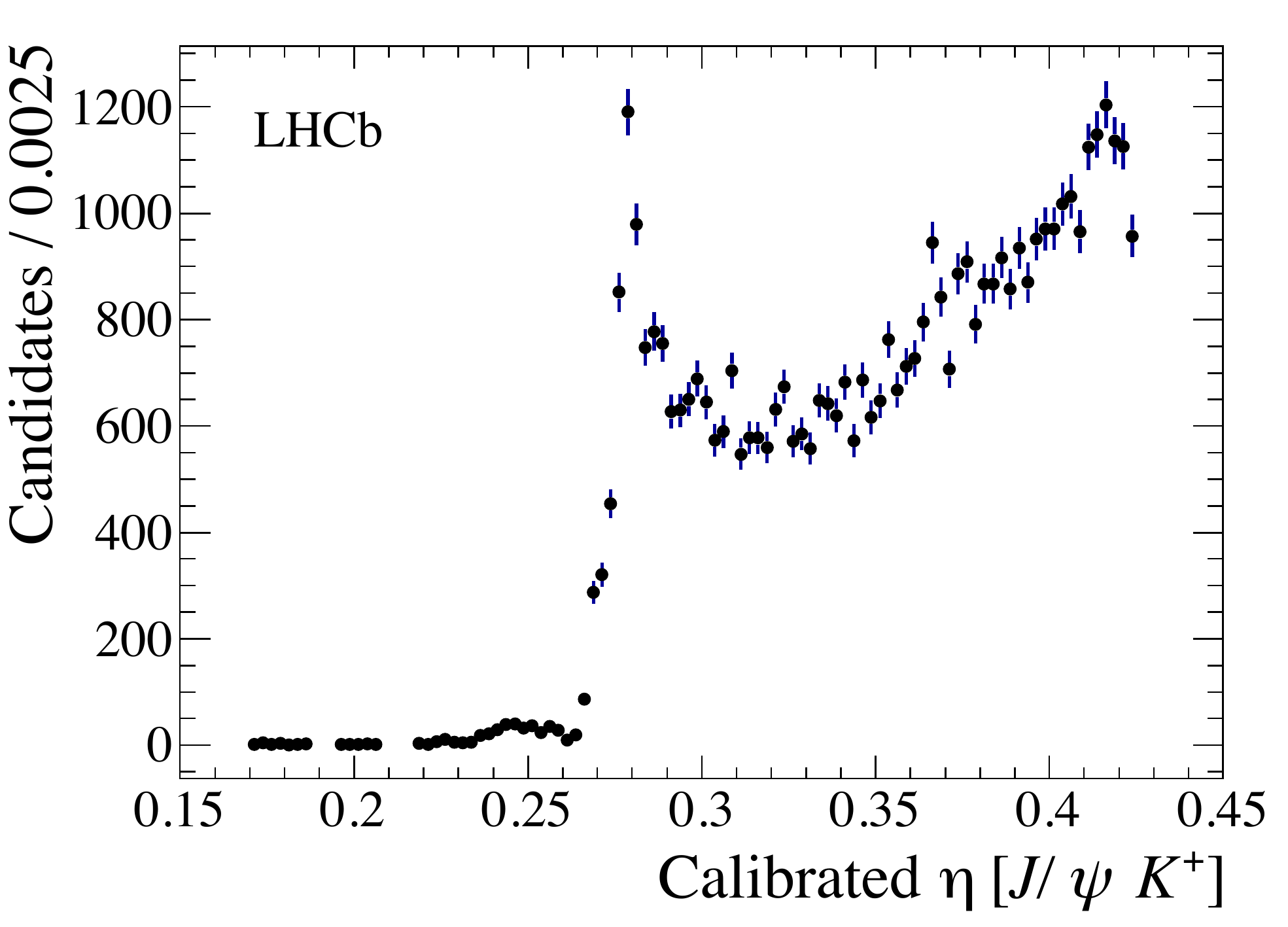}}
{\includegraphics[scale=0.5]{etadist_bjpsik.eps}}
\caption{Distribution of the calibrated predicted mistag probability $\omega(\eta)$ for the \BToJPsiK data sample.}
\label{fig:eta_data}
\end{figure}

The distribution of $\eta$ after calibration for the \BToJPsiK control sample is shown in Fig.~\ref{fig:eta_data}. 
The tagging efficiency, mistag fraction, and the tagging power of the charm tagger are reported in Table~\ref{table:power} for the training sample of simulated $\B \to \jpsi X$ decays and for both calibration channels.
The propagated statistical uncertainty of the calibration parameters dominates the statistical uncertainty of the tagging power.
The overall tagging power is slightly higher in simulation than in data, due to differences in the distributions of input variables.
The tagging powers in the two $\B \to \jpsi X$ calibration channels are consistent. 

\begin{table}
\caption[Tagging efficiencies]{Tagging efficiencies ($\etag$), effective mistag fractions ($\mistag$), and tagging powers ($\effeff$) in the various data samples studied. The first uncertainties are statistical and the second are systematic. The sample labeled Simulation is the training sample of simulated \BToJPsiK, \BdToJPsiKst, and \BsToJPsiPhi decays, which has negligible statistical uncertainties.}
\label{table:power}
\centering
\begin{tabular}{llll}
\toprule
Sample & \multicolumn{1}{c}{$\etag$} & \multicolumn{1}{c}{$\mistag$} & \multicolumn{1}{c}{$\effeff$} \\
\hline
Simulation   & \RateMCShort{}   & \MistagMCShort{}   & \PowerMCShort{} \\
\BToJPsiK & \RateXAllShort{} & \MistagXAllShort{} & \PowerXAllShort{} \\
\BdToJPsiKst & \KstRateXAllShort{} & \KstMistagXAllShort{} & \KstPowerXAllShort{} \\
\BdToDpi & \DRateXAllShort{} & \DMistagXAllShort{} & \DPowerXAllShort{} \\
\BsToDspi & \DsRateXAllShort{} & \DsMistagXAllShort{} & \DsPowerXAllShort{} \\
\bottomrule
\end{tabular}
\end{table}

Table~\ref{table:power} also reports the tagging metrics for the decays \BdToDpi and \BsToDspi.
Fits to the mass distributions of the signal candidates are performed to separate signal from background.
In each fit, the signal is modeled by a sum of two Crystal Ball functions and the combinatorial background is described by an exponential function.
Several fully and partially reconstructed backgrounds are also modeled in the fit to the \BsToDspi sample. 
The tagging efficiency for these samples is found to be higher than for the samples of $\B \to \jpsi X$ decays, due to correlations between the kinematics of the signal \B and the opposite-side charm hadrons. 
The effective mistag fraction for these samples is consistent with that on the $\B \to \jpsi X$ samples.
The net effect is an increased tagging power for these $\B \to \D X$ decays, similar to that observed for other opposite-side tagging algorithms~\cite{LHCb-PAPER-2014-059,LHCb-PAPER-2014-051}.

To use the charm tagger in a physics analysis, the flavour tagging information from the charm tagger can be combined with information from other tagging algorithms.
Assessing the actual gain in performance depends on the method of combination and calibration, as well as on the set of tagging algorithms being combined.
Due to correlations with other tagging algorithms, in particular the OS kaon and vertex-charge taggers, the maximum possible increase in tagging power after the addition of the charm tagging algorithm is less than its individual tagging power.
The performance of the combination of the current OS tagging algorithms with and without the addition of the charm tagger has been measured on the \BToJPsiK data sample.
The absolute net gain in tagging power using the current combination algorithm is found to be around $0.11\%$, compared to the current total OS tagging power of about $2.5\%$~\cite{LHCb-PAPER-2014-058,LHCb-PAPER-2014-059,LHCb-PAPER-2015-004,LHCb-PAPER-2015-005}.

\section{Conclusion}
\label{sec:conclusions}

An algorithm has been developed that determines the flavour of a signal \bquark hadron at production time by reconstructing opposite-side charm hadrons from a number of decay channels.
The flavour tagger uses boosted decision tree algorithms trained on simulated data, and has been calibrated and evaluated on data using the self-tagged decay $\Bp \to \jpsi \Kp$.
Its tagging power for data in this channel is found to be \PowerXAllLabelShort{}.
The calibration has been cross-checked using the decay \BdToJPsiKst, giving consistent results.
The tagging power is found to be higher for the decays \BdToDpi and \BsToDspi, at \DPowerXAllLabelShort{} and \DsPowerXAllLabelShort{}, respectively.

\section*{Acknowledgements}

We express our gratitude to our colleagues in the CERN
accelerator departments for the excellent performance of the LHC. We
thank the technical and administrative staff at the LHCb
institutes. We acknowledge support from CERN and from the national
agencies: CAPES, CNPq, FAPERJ and FINEP (Brazil); NSFC (China);
CNRS/IN2P3 (France); BMBF, DFG, HGF and MPG (Germany); INFN (Italy); 
FOM and NWO (The Netherlands); MNiSW and NCN (Poland); MEN/IFA (Romania); 
MinES and FANO (Russia); MinECo (Spain); SNSF and SER (Switzerland); 
NASU (Ukraine); STFC (United Kingdom); NSF (USA).
The Tier1 computing centres are supported by IN2P3 (France), KIT and BMBF 
(Germany), INFN (Italy), NWO and SURF (The Netherlands), PIC (Spain), GridPP 
(United Kingdom).
We are indebted to the communities behind the multiple open 
source software packages on which we depend. We are also thankful for the 
computing resources and the access to software R\&D tools provided by Yandex LLC (Russia).
Individual groups or members have received support from 
EPLANET, Marie Sk\l{}odowska-Curie Actions and ERC (European Union), 
Conseil g\'{e}n\'{e}ral de Haute-Savoie, Labex ENIGMASS and OCEVU, 
R\'{e}gion Auvergne (France), RFBR (Russia), XuntaGal and GENCAT (Spain), Royal Society and Royal
Commission for the Exhibition of 1851 (United Kingdom).

\addcontentsline{toc}{section}{References}
\setboolean{inbibliography}{true}
\bibliographystyle{LHCb}
\bibliography{main,LHCb-PAPER,LHCb-CONF,LHCb-DP,LHCb-TDR,my_refs}

\newpage

\centerline{\large\bf LHCb collaboration}
\begin{flushleft}
\small
R.~Aaij$^{38}$, 
B.~Adeva$^{37}$, 
M.~Adinolfi$^{46}$, 
A.~Affolder$^{52}$, 
Z.~Ajaltouni$^{5}$, 
S.~Akar$^{6}$, 
J.~Albrecht$^{9}$, 
F.~Alessio$^{38}$, 
M.~Alexander$^{51}$, 
S.~Ali$^{41}$
% ...

{\footnotesize \it
% ...
$ ^{5}$Clermont Universit\'{e}, Universit\'{e} Blaise Pascal, CNRS/IN2P3, LPC, Clermont-Ferrand, France\\
$ ^{6}$CPPM, Aix-Marseille Universit\'{e}, CNRS/IN2P3, Marseille, France\\
% ...
$ ^{9}$Fakult\"{a}t Physik, Technische Universit\"{a}t Dortmund, Dortmund, Germany\\
% ...
$ ^{37}$Universidad de Santiago de Compostela, Santiago de Compostela, Spain\\
$ ^{38}$European Organization for Nuclear Research (CERN), Geneva, Switzerland\\
% ...
$ ^{41}$Nikhef National Institute for Subatomic Physics, Amsterdam, The Netherlands\\
% ...
$ ^{46}$H.H. Wills Physics Laboratory, University of Bristol, Bristol, United Kingdom\\
% ...
$ ^{51}$School of Physics and Astronomy, University of Glasgow, Glasgow, United Kingdom\\
$ ^{52}$Oliver Lodge Laboratory, University of Liverpool, Liverpool, United Kingdom\\
% ...
}
\end{flushleft}

\end{document}